# Cloud Computing: Applications, Challenges and Open Issues

▶ Sahil Mishra
B. Tech. Student, IIITDM Kurnool, Andhra Pradesh

▶ Sanjaya Kumar Panda
Assistant Professor and Head (CSE), IIITDM Kurnool, Andhra Pradesh
Email: sanjayauce@gmail.com

**Background:**

Cloud computing is one of the innovative computing, which deals with storing and accessing data and programs over the Internet [1]. It is the delivery of computing resources and services, such as storing of data on servers and databases, providing networking facilities and software development platforms over the Internet. It provides the flexibility of resources for everyone. These services are provided via data centers, which are located in various parts of the world [2, 3]. Cloud computing makes access to these resources to everyone on a global scale at a very minimal cost and significantly higher speed. These servers provide services to the users, which would have cost a lot of computational power to them if they had to buy them. The first mention of cloud computing was referenced in a Compaq internal document released in 1996 [4]. Cloud computing was then commercialized in 2006 when Amazon released elastic compute cloud (EC2). Furthermore, Google released Google app engine in 2008 and Microsoft Azure services were launched in October 2008, which increased the competition in the area of cloud computing. Since then these companies have done a lot of development in cloud computing.

**Types and Services:**

All the servers or clouds are not identical and one of the clouds is not fulfilling the requirements for everyone. As a result, there are various architectures and models of clouds, which serve the specific purpose based upon the needs of the users. There are three ways to run cloud services, namely public cloud, private cloud and hybrid cloud. The third-party cloud service providers own the public cloud and they provide the users with the storage services on servers and databases over the Internet. Moreover, they provide a lot of other services like virtual machines for processing large amounts of data and programs, which is not possible to be processed on a single computer [5]. Private cloud, as the name suggests, refers to the resources provided to a single organization, located locally in its premise or sometimes hosted on the off premise. It is usually used by organizations due to the sensitivity of the data and other privacy concerns. On the other hand, hybrid cloud is the mixture of resources of public cloud and private cloud by permitting the data and programs to be shared between them. It plays an important role when the computing and processing demand fluctuates [6]. It allows the seamless scale-up of resources to a public cloud to handle the fluctuating demand, without giving third-party access to the entire data. The sensitive data lies safely behind the firewall. These clouds offer various types of services in which infrastructure as a service (IaaS) is one of them. It offers services that include remote access of a full-pledged computer infrastructure, like virtual computers, servers, storage devices, etc. The other service offered by the cloud is platform as a service (PaaS), which allows user to host the application on remote server. It provides required resources, such as middleware, database, operating system, networking and many more. The third type of service is software as a service (SaaS), which provides the applications to the users via the Internet.

**Applications in Society:**

Cloud computing is used in various sectors of society. Government organizations use the cloud to store the data of various departments. They make the use of IaaS to store data on the servers. They even host applications on the cloud using PaaS. Some organizations use the private cloud due to sensitive data [7]. Cloud computing has made it easy for the government organizations to share data and collaborate among various departments. National informatics centre (NIC) offers the cloud services to various government organizations in India. It provides a private cloud to handle the jobs of Indian government organizations. It is responsible for hosting websites and handling data of various examinations conducting councils and ministries. Multiple business organizations and industries make use of PaaS and SaaS to host their applications on the Internet and make them available to the users. SaaS is used by Google to provide various applications to users over the Internet, such as Google documents, sheets, slides and forms. Google also provides a Google cloud platform (GCP), which offers all the services (i.e., IaaS, PaaS and SaaS). GCP is used by many software developers to build and deploy applications at a very low cost. It is even used to collaborate with other users and it also offers them the flexibility to expand according to their needs at a very low cost. Various companies and organizations also use GCP for training machine learning and deep learning models.

**Challenges and Open Issues:**

In the month of January 2018, RightScale initiated its annual survey on the trends of cloud [8]. They interviewed 997 technical professionals working in different organizations using the cloud. They found various challenges cloud, which are presented as follows. Security and privacy have always been a concern for cloud computing. The third-party service providers can access the data of users and may sell it to other organizations, without any permission from the user. The data can be misused for many purposes like identifying personality trait of a particular user and its political interests of the user. According to the cloud security alliance, the main threats posed to data. This is due to insecure interfaces and application programming interface (API), data loss and leakage, and hardware failure [4]. The other privacy concern is the hacking. If the hackers hack the data servers, they would be getting a handful of the sensitive data of various users across the globe. As a result, as far as privacy concerns, some government and private organizations prefer using the private cloud. It provides physical control of data and more security than keeping it in the public cloud.

Cloud computing is very cost efficient for the users as well as the organizations. They can easily expand their processing capabilities, without spending large amounts of money in the hardware. However, on-demand availability and scalability of cloud computing make it harder to predict the required quantity in future [5]. Even the server failures due to lack of maintenance incur huge losses. The other big challenge cloud computing faces are the lack of resources like data centers and cloud engineers. As the time is changing, the users of cloud computing are increasing at an exponential rate, which may result in failure of data centers. This puts a lot of pressure on the service providers to get enough resources for the huge number of users. This also incurs a lot of cost to them. Even the power consumption by these data centers is quite high. Therefore, service providers like Microsoft are trying to reduce it by putting the data centers in the ocean. This reduces the cost of cooling them to a large extent. Cloud servers handle a lot of data and computations. As a circumstance, they need continuous monitoring and supervision. If any technical glitch occurs on the server, then a lot of users face its consequences, which results in loss of time, data and money to both users and service providers. Technical faults are also sometimes caused if the consumer, especially the big organizations does not implement it properly. This not only costs them unnecessary overhead, but also the servers to handle vague data and process them, which may again result in some technical faults in the servers.

Cloud computing has served as a boon to the mankind. Users can store the data in the cloud and access it anytime and anywhere, but accessing them requires the Internet connectivity. In order to access large amounts of data, good internet connectivity is required, which is not same at every part of the world. Therefore, various places are refraining from using the cloud services. Sometimes, the users and organizations tend to change the service providers due to a lot of issues. Therefore, ensuring resource portability is very necessary. Cloud technology must have the capability to transfer and integrate resources on other servers, without any necessary overhead. If a user deploys an application on a server, then migrating it to the server of another service provider requires the user to modify the application according to the requirement of new server. Even the user cannot share resources between the servers of different service providers.

**Conclusion:**

Cloud computing has opened the gates for large storage and computational power at a very minimal cost. Consumers don't need to buy expensive resources to carry on their daily jobs. But, in spite of having a lot of pros, the cloud computing also has a significant number of cons. These issues have always been prevalent in the field of cloud computing. Various attempts are being made to reduce these challenges to a certain level by the people of different domains. Despite of facing these challenges, cloud computing is still helping the technology to reach to the people at all the remote places across the globe.

## About the Authors

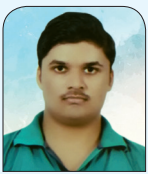
**Mr. Sahil Mishra** is currently pursing B. Tech. degree from IIITDM Kurnool, Andhra Pradesh, India. He has published few papers in reputed international journals. His research interests include cloud computing, recommender systems and big data analytics.

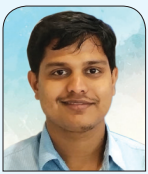
**Dr. Sanjaya Kumar Panda** (I1504530) is working as an Assistant Professor and Head in the Department of CSE at IIITDM Kurnool, Andhra Pradesh, India. He received Ph. D. degree from IIT (ISM) Dhanbad, Jharkhand, India and M. Tech. degree from NIT Rourkela, Odisha, India in CSE. He received two silver medal awards for best graduate and best post-graduate in CSE. He also received CSI Young IT Professional Award, CSI Paper Presenter Award at International Conference and CSI Distinguished Speaker Award. He has published more than 60 papers in reputed journals and conferences. His research interests include cloud computing, recommender systems and big data analytics.